# Compared study of Shannon, Tsallis and Gaussian entropy of bound magnetopolaron in nanostructures

M. Tiotsop[1*], A. J. Fotue[1†], H. B. Fotsin[2], L. C. Fai[1]

[1] *Mesoscopic and Multilayers Structures Laboratory, Department of Physics, Faculty of Science, University of Dschang, P.O. Box 479 Dschang, Cameroon*

[2] *Laboratory of Electronics and Signal Processing, Department of Physics, Faculty of Science, University of Dschang, P.O. Box 67 Dschang, Cameroon*

## Abstract

Many methods have been experimented to study decoherence in nanostructures. Tsallis, Shannon and Gaussian entropy have been used to study decoherence separately; in this paper, we compared the results of the sus-mentioned entropies in nanostructures. The linear combination operator and the unitary transformation was used to derive the magnetopolaron spectrum that strongly interact with the LO phonons in the presence of electric field in the pseudo harmonic and delta quantum dot. Numerical results revealed for the quantum pseudo dot that: *(i)* The amplitude of Gauss entropy is greater than the amplitude of Tsallis entropy which inturn is greater than the amplitude of Shannon entropy. The Tsallis entropy is not more significant in nanostructure compared to Shannon and Gauss entropies, *(ii)* With an increase of the zero point, the dominance of the Gauss entropy on the Shannon entropy was observed on one hand and the dominance of the Shannon entropy on the Tsallis entropy on the other hand ; this suggested that in nanostructures, Gauss entropy is more suitable in the evaluation of the average of information in the system, for the delta quantum dot it was observed that *(iii)* when the Gauss entropy is considered, a lot of information about the system is missed. The collapse revival phenomenon in Shannon entropy was observed in RbCl and GaAs delta quantum dot with the enhancement of delta parameter; with an increase in this parameter, the system in the case of CsI evolved coherently; with Shannon and Tsallis entropies , information in the system is faster and coherently exchanged; *(iv)* The Shannon entropy is more significant because its amplitude outweighs on the others when the delta dimension length enhances. The Tsallis entropy involves as wave bundle; which oscillate periodically with an increase of the oscillation period when delta dimension length is improved.

---

[*] maurice.tiotsop@univ-dschang.org
[†] fotuea@yahoo.fr

# I. INTRODUCTION

Nanoscience studies the novel phenomena and properties of materials that occur on the nanoscale that is the size of atoms and molecules. The field of nanotechnology is dedicated to the application of nanoscience to produce new materials and devices. In these materials, quantum effects such as discrete energy levels due to size quantization, shell filling and various spin effects of electrons confined in so-called quantum dots have been observed. The fundamental quantum mechanics for these types of artificial atoms is the same as for natural atoms. However, there are several important differences that make these systems an excellent laboratory to investigate various aspects of quantum mechanics and give us access to exciting new regimes that are impossible to achieve in natural atoms. These systems are therefore very suitable for fundamental scientific study, but also offer great technological promise, as witnessed by the emerging fields of coherent electronics, spintronics, quantum computation and information.

Since the introduction of this last field by Paul Benioff [1], Richard Feynman [2] and David Deutsch [3], it has received a lot of attention and is still a hot topic in nanoscience. Quantum computers use the non-locality of quantum physics to allow exponentially fast solutions to classical problems [4]. The fundamental element of a quantum computer is the quantum bit (qubit), which may be in a superposition state of zero and one. It is a very frail state. Ideally, the quantum computer is a closed system, but in reality, when information leaks out, the qubits collapse and errors are introduced into the calculation. Evaluation of the impact of this decoherence process is a key to understanding the feasibility of quantum computation [5-7]. Efforts have been done to control the decoherence in nanostructures [8-13].

The very promising aspect of the quantum research is the using the state of the quantum dot as a qubit in quantum computing. Several schemes, like trapped ions [14] quantum optical systems [15] nuclear and electron spins [16-18] and superconductor Josephson junctions have been proposed for realizing quantum computation [19-22]. Muhonen and cowokers [23] showed the application of several concepts and tools of weak single-shot measurements to a model solid-state spin system. They demonstrated the ability to coherently control a nuclear spin using only electron spin resonance pulses and electron spin readout, and how to measure tunnel rates without any tunnelling events. Cai et *al*. [24] investigated the influence of temperature and magnetic field on the first excited state of a CsI quantum pseudodot. Muhonen et *al*. [25] presented the coherent operation of individual $^{31}$P electron and nuclear spin qubits in a top-gated nanostructure, fabricated on an isotopically engineered $^{28}$Si substrate. The $^{31}$P nuclear spin sets the new benchmark coherence time of any single qubit in the solid state and reaches >99.99% control fidelity.

In the theory of information, entropy is the average information in a considered state; it also measures the degree of disorder of a considered system. It is useful in the study of information storage [26-27]. The more suitable and most popular entropy is the Shannon entropy which is the amount of transmitted message as introduced by Claude Shannon in 1948 [28]. The Shannon entropy is a decreasing function of the scattering of random variable, and is maximal when all the outcomes are equally likely. Constantino Tsallis [29] and Alfred Rényi [30] both proposed generalized entropies Of the Shannon entropy. More recently the Gaussian entropy [31] has been used to study the amount of information in a considered system. Following these work many efforts were done in the improvement theory and the study of information theory. Robert Sneddon [32] provides a solution to the encoding problem by deriving a method to estimate the Tsallis entropy in natural data. Rényi [33] in 1965 answered the fundamental question of information theory: how should the amount of information be measured? Tomasz Maszczyk and Wlodzislaw Duch [34] modified C4.5 decision trees based on Tsallis and Renyi entropies and tested on several high-dimensional microarray datasets with interesting results. The approach may be used in any decision tree and information selection algorithm. Sanei Tabass Manije et *al.* [35] presented the conditional Tsallis entropy on the basis of the conditional Rényi entropy. Regarding the fact that Rényi entropy is the monotonically increasing function of Tsallis entropy, a relationship has also been presented between the joint Tsallis entropy and conditional Tsallis entropy. Shannon and Gaussian have been used by Khordad and Sedehi [36] to study decoherence in asymmetric quantum dot. We [37] used Tsallis entropy to study the decoherence of polaron in CsI quantum dot and showed that the Tsallis entropy evolved as a wave envelop that increases with an increase of non-extensive parameter and with an increase of electric field strength, zero point of pseudo dot and cyclotron frequency, the wave envelop evolved periodically with reduction of period. The decoherence of polaron in RbCl triangular quantum was studied in [38].

This article aims at the comparison of Tsallis, Shannon and Gaussian entropies in the study of decoherence of magnetopolaron in various nanostructures as RbCl, CsI and GaAs quantum dot. It has the following structure: In section 2, we describe the Hamiltonian of the system and used the results in [38, 39] to derive the Tsallis, Shannon and Gaussian entropy in polar crystal. In section 3, we discussed obtained results and ended section 4 with a conclusion.

## II. MODEL AND CALCULATION

An electron moving in nanostructure quantum dot with a three-dimensional potential and interacting with bulk LO phonons under the influence of an electric and a magnetic field is considered. The electric field $F$ is directed along the $\rho-$direction and the magnetic field is along the $z-$direction with vector potential $A = B(-y/2, -x/2, 0)$. Due to the presence of the magnetic field, the entity resulting from the interaction of electron and LO phonons is called magnetopolaron. To study this entity, we considered the following Hamiltonian:

$$H = \frac{1}{2m}\left(p_x - \frac{\beta^2}{4}y\right)^2 + \frac{1}{2m}\left(p_y + \frac{\beta^2}{4}x\right)^2 - e^*\rho F + \frac{p_z^2}{2m} + V(r)$$
$$+ \sum_q \hbar\omega_{LO} a_q^+ a_q + \sum_q \left[V_q a_q \exp(iq.r) + h.c.\right] \quad (1)$$

The potential $V(r)$ takes different according to the configuration and is defined as follow:

Pseudo harmonic potential,

$$V(r) = V_0\left(\frac{r}{r_0} - \frac{r_0}{r}\right)^2, \quad (2)$$

Delta potential

$$V(z) = -\frac{\hbar^2 \kappa}{2ma}\delta(z), \quad (3)$$

where the meanings of the physical quantities in eqs. 1-3 are defined in [39]. Let recall our results obtained for the ground and first excited state in [38]

Pseudo harmonic potential

$$E_0 = \frac{\hbar^2 \lambda_0}{2m} + \frac{\hbar^2 \mu_0}{4m} + \frac{m\omega_c^2}{8\lambda_0} - \frac{\sqrt{\pi}e^* F}{2\sqrt{\mu_0}} + V_0\left(\frac{1}{\lambda_0 r_0^2} + \lambda_0 r_0^2 - 2\right) -$$
$$- 8\hbar\alpha\sqrt{\frac{\hbar\omega_{LO}}{2m}}\sqrt{2\lambda_0/\pi\left(1 - \frac{\lambda_0}{\mu_0}\right)}\arcsin\left(1 - \frac{\lambda_0}{\mu_0}\right)^{1/2} \quad (4)$$

and

$$E_1 = \frac{\hbar^2 \lambda_1}{2m} + \frac{3\hbar^2 \mu_1}{4m} + \frac{m\omega_c^2}{8\lambda_1} - \frac{\sqrt{\pi} e^* F}{2\sqrt{\mu_1}} + V_0 \left( \frac{1}{r_0^2} + \lambda_1^2 r_0^2 - 2\lambda_1 \right)$$
$$- 8\hbar\alpha \sqrt{\frac{\hbar\omega_{LO}}{2m}} \sqrt{2\lambda_1 / \pi \left(1 - \frac{\lambda_1}{\mu_1}\right)} \arcsin\left(1 - \frac{\lambda_1}{\mu_1}\right)^{1/2} \quad (5)$$

Delta potential

$$E_0 = \frac{\hbar^2 \lambda_0}{2m} + \frac{\hbar^2 \mu_0}{4m} + \frac{m\omega_c^2}{8\lambda_0} - \frac{\sqrt{\pi} e^* F}{2\sqrt{\mu_0}} - \frac{K}{2ma^2} \sqrt{\frac{\lambda_0}{\pi}} -$$
$$- 8\hbar\alpha \sqrt{\frac{\hbar\omega_{LO}}{2m}} \sqrt{2\lambda_0 / \pi \left(1 - \frac{\lambda_0}{\mu_0}\right)} \arcsin\left(1 - \frac{\lambda_0}{\mu_0}\right)^{1/2} \quad (6)$$

and

$$E_1 = \frac{\hbar^2 \lambda_1}{2m} + \frac{3\hbar^2 \mu_1}{4m} + \frac{m\omega_c^2}{8\lambda_1} - \frac{\sqrt{\pi} e^* F}{2\sqrt{\mu_1}} - \frac{K}{2ma^2} \sqrt{\frac{\lambda_1}{\pi}} -$$
$$- 8\hbar\alpha \sqrt{\frac{\hbar\omega_{LO}}{2m}} \sqrt{2\lambda_1 / \pi \left(1 - \frac{\lambda_1}{\mu_1}\right)} \arcsin\left(1 - \frac{\lambda_1}{\mu_1}\right)^{1/2} \quad (7)$$

The time evolution of the state of the electron can then be written as

$$\psi_{01}(t,r) = \frac{1}{\sqrt{2}} |0\rangle \exp\left(-i\frac{E_0 t}{\hbar}\right) + \frac{1}{\sqrt{2}} |1\rangle \exp\left(-i\frac{E_1 t}{\hbar}\right) \quad (8)$$

The probability density is in the following form:

$$Q(t,r) = |\psi_{01}(t,r)|^2 \quad (9)$$

Constantino Tsallis[29] proposed the new entropy formalism and generalise the entropy as

$$S_q = k \frac{1 - \sum_{i=1}^{w} Q_i^q}{1 - q} \quad (10)$$

where $q$ is a real number, $k$ is a conventional positive constant and $\sum_{i=1}^{w} Q_i = 1$.

Equation(10) can take the form

$$S_q = -k \frac{\sum_{i=1}^{w} Q_i^q}{q-1} + \frac{\sum_{i=1}^{w} Q_i}{q-1} = k \left( \sum_{i=1}^{w} \frac{Q_i - Q_i^q}{q-1} \right) \qquad (11)$$

With the continuous wave function (11) takes the following forms

$$S_q(t) = k \int_0^\infty dr \left( \frac{Q - Q^q}{q-1} \right) \qquad (12)$$

Applying the replica-trick type of expansion, he showed that[29, 37]

$$S_1 \equiv \lim_{q \to 1} S_q = k \lim_{q \to 1} \frac{1 - \sum_{i=1}^{w} Q_i \exp[(q-1)\ln Q_i]}{q-1} = -k \sum_{i=1}^{w} Q_i \ln Q_i \qquad (13)$$

which is the well-known Shannon entropy[28]. The non-extensive entropy with Gaussian gain defined Susan and Hanmandlu [40] is given as

$$S(t) = \sum_{i=1}^{w} Q_i \exp(Q_i^2) \qquad (14)$$

In its continuous form, it is written as

$$S(t) = \int_0^\infty dr\, Q(r,t) \exp(Q^2(r,t)) \qquad (15)$$

After describing the spectrum of bound magnetopolaron in nanostructures and various entropy, the next part will be dedicated to the results.

## III. RESULTS

Numerical results of the Shannon, Gaussian and Tsallis entropy versus time for different various values chemical potential $V_0$, zero point of pseudo dot $r_0$, delta parameter $\kappa$ and delta dot length $a$. Numerical calculations for GaAs, RbCl and CsI crystals were performed, using experimental parameter values defined in the table:

| Materials | GaAs[a] | RbCl[b] | CsI[c] |
|---|---|---|---|
| $\hbar\omega_{LO}$ | 36.4 meV | 21.639 meV | 11.0 meV |
| $m/m_0$ | 0.067 | 0.432 | 0.42 |
| $\alpha$ | 0.2 | 3.81 | 3.67 |

[a] Reference [41]
[b] Reference [42]
[c] Reference [43]

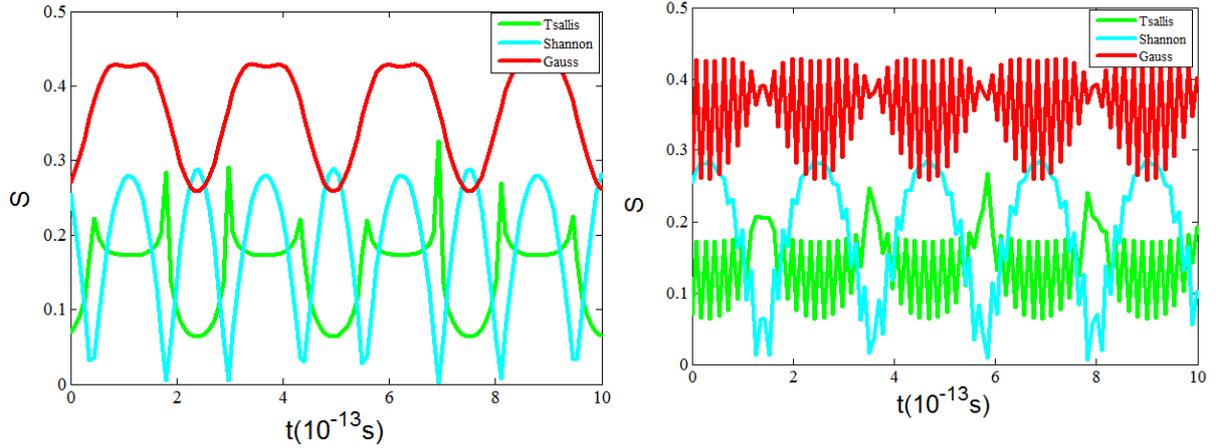

FIG. 1. Entropy versus time for magnetopolaron in GaAs pseudo harmonic potential with $r_0 = 5nm$, $\omega_c = 5 \times 10^{-13} s^{-1}$, $F = 10V/nm$ and $q = 0.05$ (a) $V_0 = 5meV$ and (b) $V_0 = 100meV$.

Figs. 1-3 show the time evolution of various type of entropy of magnetopolaron in pseudo harmonic potential for different values of chemical potential. The amplitude of Gauss entropy is greater than the amplitude of Tsallis entropy which is greater than the amplitude of Shannon entropy. With the enhancement of the chemical potential for the polaron in GaAs and RbCl dot, the Tsallis, Shannon and Gauss entropies have three behaviors; the increase , the constant and decrease phasis. In the theory of information, the increase phasis is synonym to the decrease of signal intensity and the decrease phasis means the gain of information about the system; the constant behavior is the normal transmission throughout the system. For Figs. 1a and 2b, the period of oscillation is $208fs$ for Gauss entropy, $108fs$ for Shannon entropy and $205fs$. From where it follows that with the Shannon entropy, the information is quickly transmitted through the system; with the Gauss entropy the information about the system is optimise. With the vanishing of Shannon entropy in GaAS, RbCl and CsI quantum pseudo dot at a certain time , it is sure that at the precise time, the information is transmitted faithfully. This plot also shows that, the Tsallis entropy is not more significant in nanostructure compared to Shannon and Gauss entropies. In Figs. 1b and 2a we observed the oscillatory behaviour of the entropy under a standing wave envelope, giving rise to the phenomenon of collapse revival in the entropy and meaning the coherent exchange of energies in the system.

Figs. 4-6 display the time evolution of Gauss, Shannon and Tsallis entropy in GaAs, RbCl and CsI pseudo harmonic potential respectively. In terms of amplitude results clearly show the dominance of the Gauss entropy on the Shannon entropy on one hand and the dominance of the Shannon entropy on the Tsallis entropy on the other hand ; this suggest that in nanostructures, Gauss entropy is more suitable in the evaluation of the average of information in the system. In the GaAs quantum pseudodot, when $t \geq 800 fs$ the Tsallis entropy is more pronounce that the Shannon entropy (Fig. 4a); here also with the increase of time, it is noted that the various entropies tend to be constant. Figs. 4 and 5 show that with an increase of the zero point, the oscillation pseudo period of entropies in RbCl and CsI quantum pseudo dot increases suggesting attenuation in the transmission of information through the system while in the GaAs it is an inverse scenario.

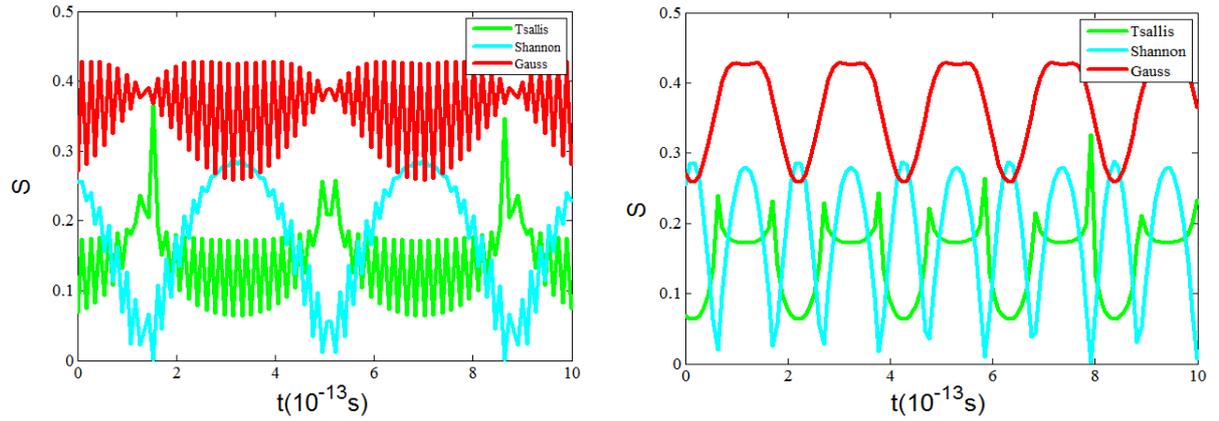

FIG. 2: Entropy versus time for magnetopolaron in RbCl pseudo harmonic potential with $r_0 = 5nm$, $\omega_c = 5 \times 10^{-13} s^{-1}$, $F = 10V/nm$ and $q = 0.05$ (a) $V_0 = 5meV$ and (b) $V_0 = 100meV$.

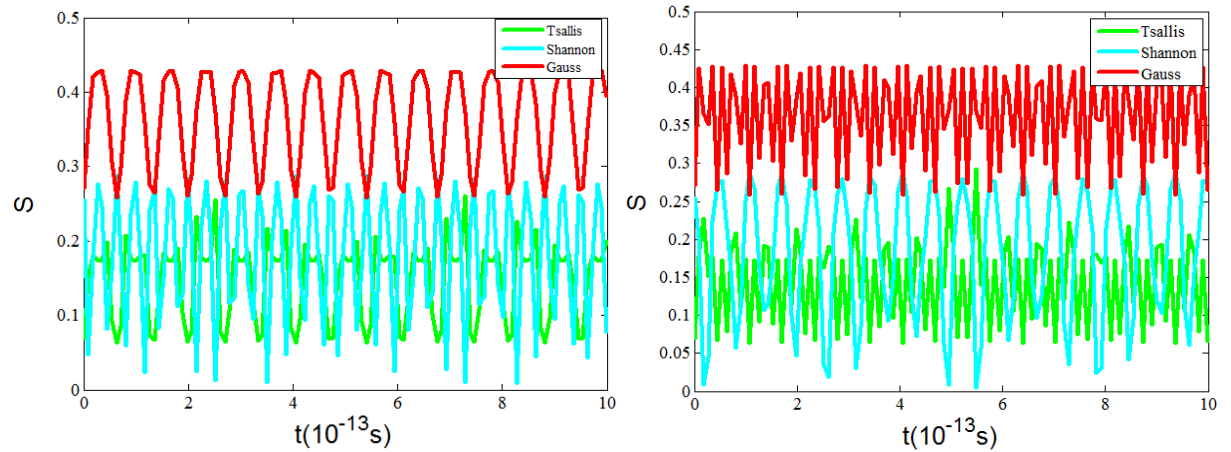

FIG. 3: Entropy versus time for magnetopolaron in CsI pseudo harmonic potential with $r_0 = 5nm$, $\omega_c = 5 \times 10^{-13} s^{-1}$, $F = 10V/nm$ and $q = 0.05$ (a) $V_0 = 5meV$ and (b) $V_0 = 100meV$

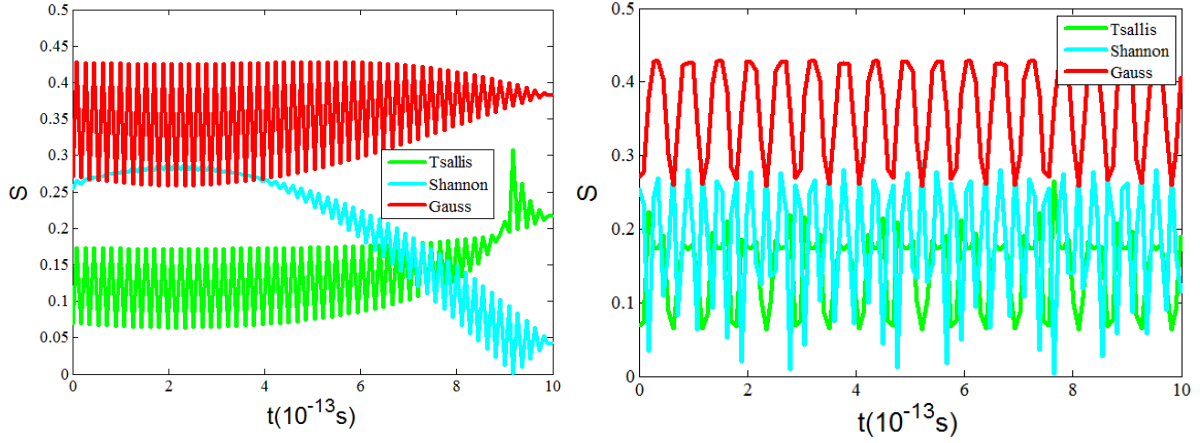

FIG.4: Entropy versus time for magnetopolaron in GaAs pseudo harmonic potential with $V_0 = 10 meV$, $\omega_c = 5 \times 10^{-13} s^{-1}$, $F = 10 V/nm$ and $q = 0.05$ (a) $r_0 = 10 nm$ and (b) $r_0 = 20 nm$

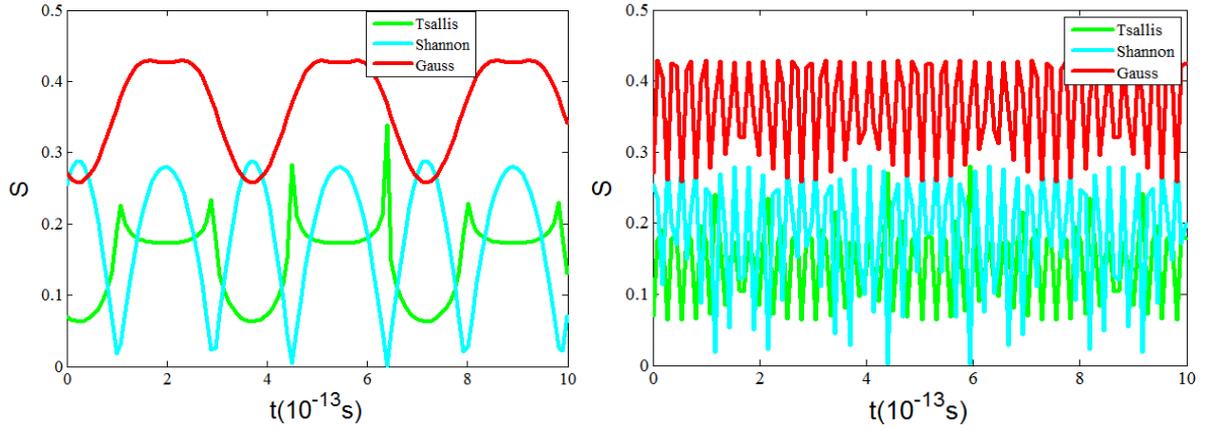

FIG. 5: Entropy versus time for magnetopolaron in RbCl pseudo harmonic potential with $V_0 = 10 meV$, $\omega_c = 5 \times 10^{-13} s^{-1}$, $F = 10 V/nm$ and $q = 0.05$ (a) $r_0 = 10 nm$ and (b) $r_0 = 20 nm$

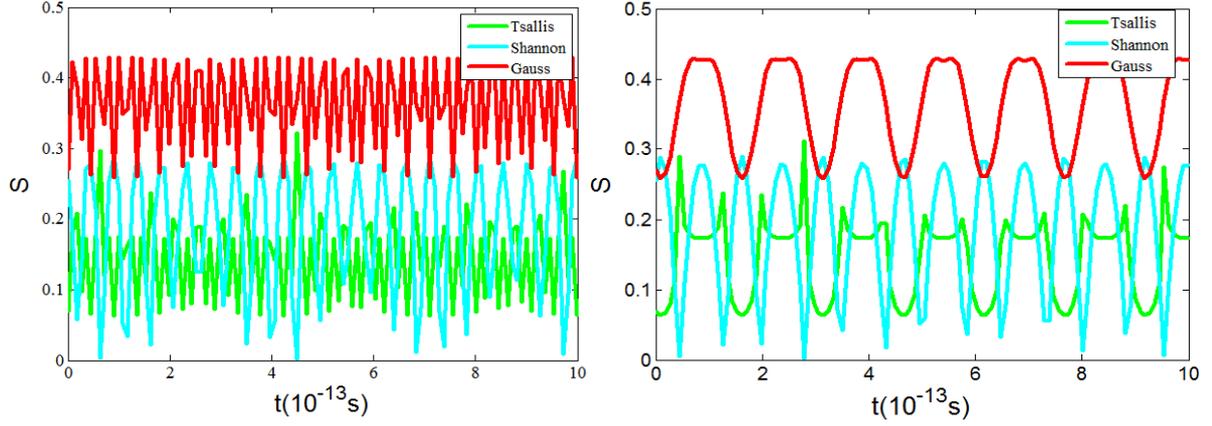

FIG. 6: Entropy versus time for magnetpolaron in CsI pseudo harmonic potential with $V_0 = 10 meV$, $\omega_c = 5 \times 10^{-13} s^{-1}$, $F = 10 V/nm$ and $q = 0.05$ (a) $r_0 = 10 nm$ and (b) $r_0 = 20 nm$

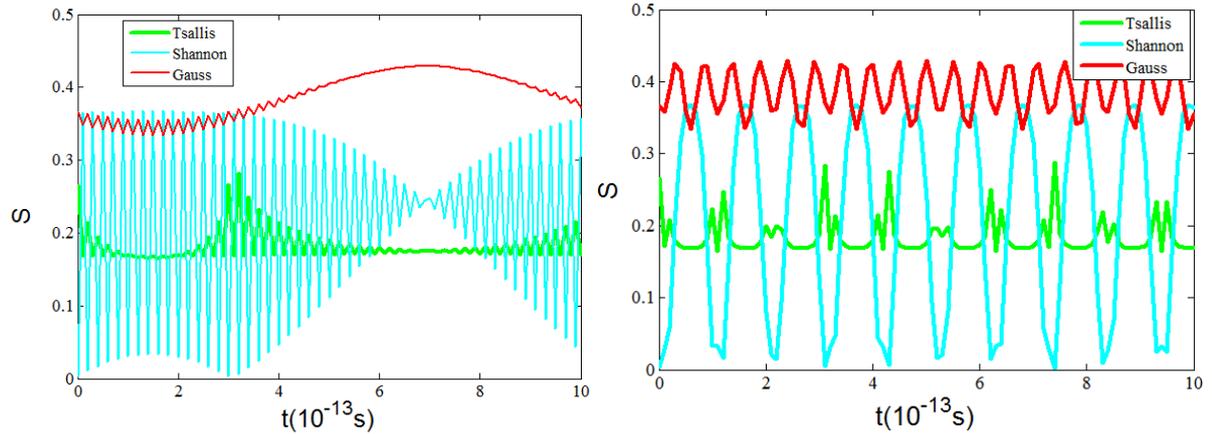

FIG. 7: Entropy versus time for magnetopolaron in GaAs delta dot with $a = 0.9$, $\omega_c = 10 \times 10^{-13} s^{-1}$, $F = 20 V/nm$ and $q = 0.05$ (a) $\kappa = 10$ and (b) $\kappa = 20$

Figs. 7 to 9 display the time evolution of various entropies in GaAs, RbCl and CsI respectively of polaron in delta quantum dot for various value of delta parameter. Entropies evolve under an enveloppe function. When the Gauss entropy is considered, a lot of information about the system is missed. The collapse revival phenomenon in Shannon entropy is observe in RbCl and GaAs delta quantum dot with the enhancement of delta parameter; with the increase of this parameter the system in the case of CsI evolve coherently; and the period of entropy increases meaning the deceleration in the information transfer process. One concluded from here that with Gauss entropy, information about the system is missed; with Shannon and Tsallis entropies , information in the system is faster and coherently exchanged.

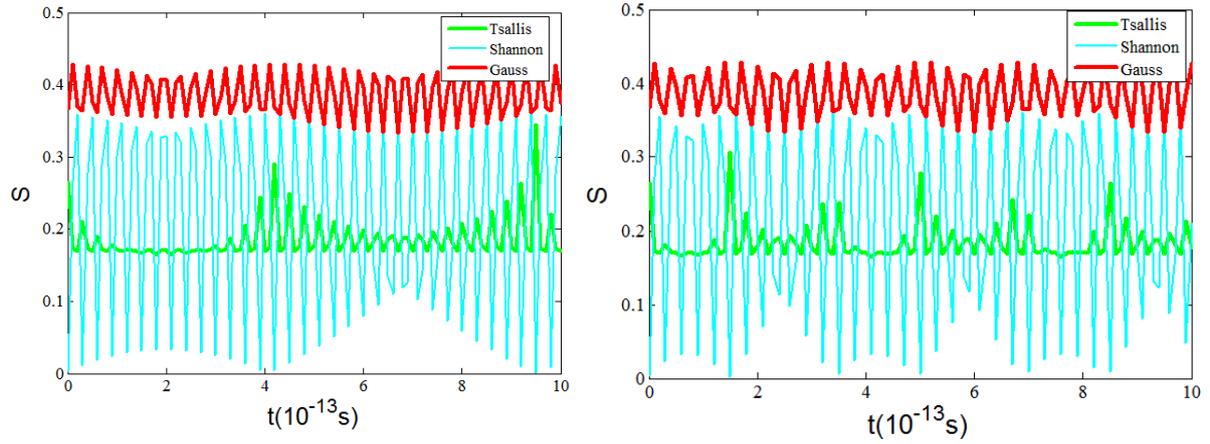

FIG. 8: Entropy versus time for magnetopolaron in RbCl delta dot with $a = 0.9$, $F = 20V/nm$ $\omega_c = 10 \times 10^{-13} s^{-1}$ and $q = 0.05$ (a) $\kappa = 10$ and (b) $\kappa = 20$

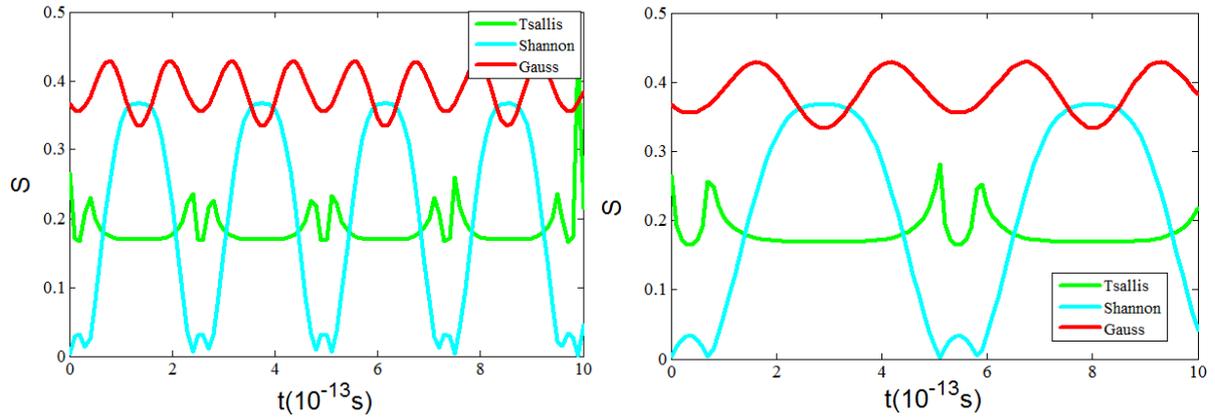

FIG. 9: Entropy versus time for magnetopolaron in CsI delta dot with $a = 0.9$, $\omega_c = 10 \times 10^{-13} s^{-1}$, $F = 20V/nm$ and $q = 0.05$ (a) $\kappa = 10$ and (b) $\kappa = 20$

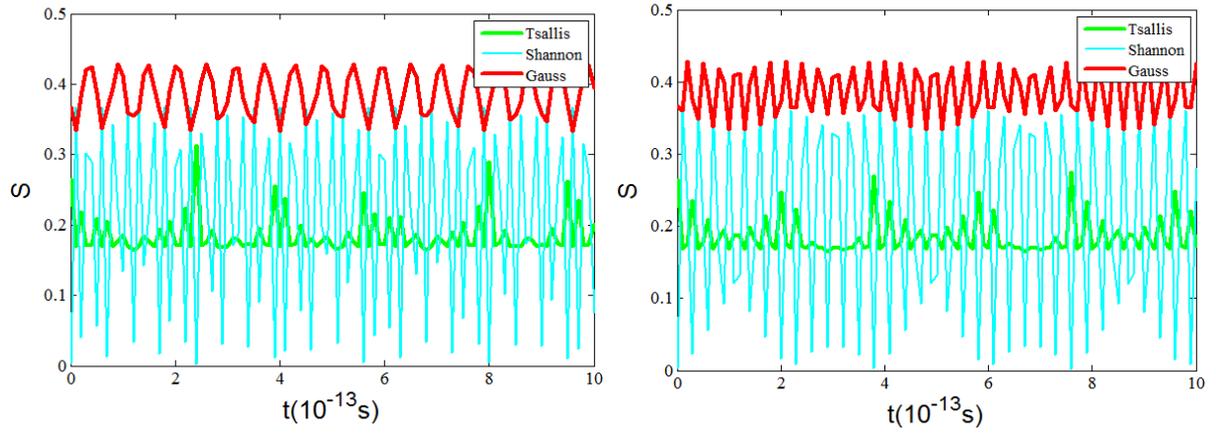

FIG. 10: Entropy versus time for magnetopolaron in GaAs delta dot with $\kappa = 30$, $\omega_c = 10 \times 10^{-13} \, s^{-1}$, $F = 20V/nm$ and $q = 0.05$ (a) $a = 0.4$ and (b) $a = 0.6$

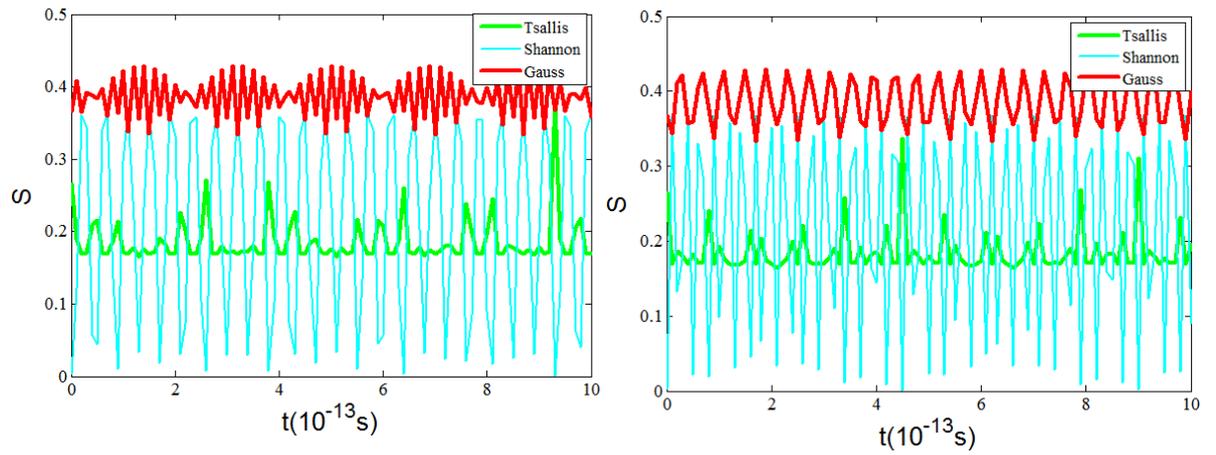

FIG.11: Entropy versus time for magnetopolaron in RbCl delta dot with $\kappa = 30$, $\omega_c = 10 \times 10^{-13} \, s^{-1}$, $F = 20V/nm$ and $q = 0.05$ (a) $a = 0.4$ and (b) $a = 0.6$

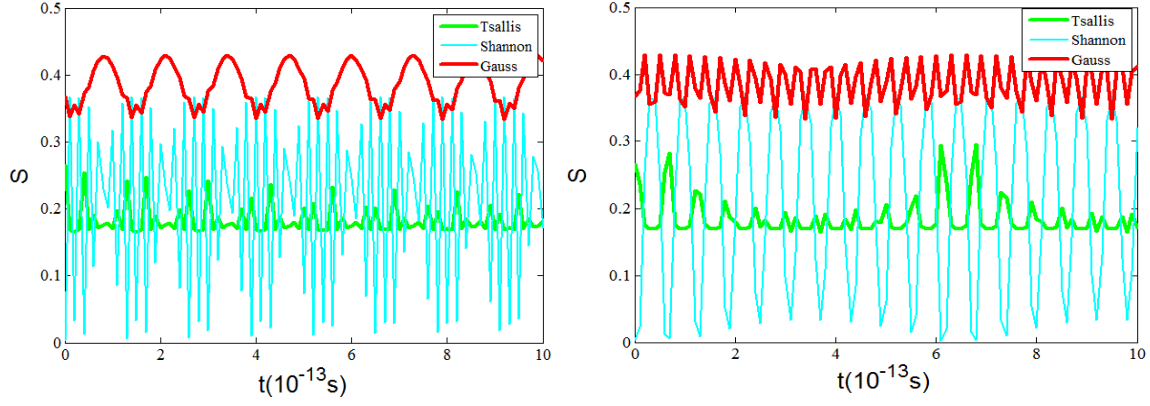

FIG. 12: Entropy versus time for magnetopolaron in CsI delta dot with $\kappa = 30$, $\omega_c = 10 \times 10^{-13}\, s^{-1}$, $F = 20V/nm$ and $q = 0.05$ (a) $a = 0.4$ and (b) $a = 0.6$

In Figs. 10 to 12, the evolution of various entropies in different nanostructures has been plotted for various value of delta dimension length $a$. The Shannon entropy here is more significant because its amplitude outweighs on the others when $a$ enhances, the period of entropy decreases for the RbCl quantum dot and increases for CsI and GaAs quantum delta dot; as we notice that the decrease of the period is synonym to the faster transmission of the information through the system. The Tsallis entropy evolves as wave bundle; which oscillate periodically with an increase of the oscillation period when $a$ is improved. Using the entropy of Tsallis, one can get the same information about the system that as the entropy of Shannon; the only difference is that the latter is more significant in terms of volume of the information carried.

## IV. CONCLUSIONS

We used the linear combination operator and unitary transformation methods to derive the ground and first excited state energy of magnetopolaron in pseudo harmonic and delta quantum dot. Three entropies (Gauss, Shannon and Tsallis) in GaAs, RbCl and CsI quantum dot have been used to study their impact on the decoherence. One can suggest the following results for application in nanotechnology: for the quantum pseudo dot, *(i)* the amplitude of Gauss entropy is greater than the amplitude of Tsallis entropy which is greater than the amplitude of Shannon entropy. With the enhancement of the chemical potential for the magnetopolaron in GaAs and RbCl dot, the Tsallis, Shannon and Gauss entropies have three behaviors; the increase , the constant and decrease phasis. In the theory of information, the increase phasis is synonym to the decrease of the signal intensity and the decrease phasis means the gain of information about the system; the constant behavior is the normal transmission throughout the system. With the Shannon entropy, the information is quickly transmitted through the system; with the Gauss entropy the information about the system is optimise. The Tsallis entropy is not more significant in nanostructure compared to Shannon and Gauss entropies. *(ii)* With the increase of the zero point, results clearly show the dominance of the Gauss entropy on the Shannon entropy on one hand and the dominance of the Shannon entropy on the Tsallis entropy on the other hand ; this suggest that in nanostructures, Gauss entropy is more suitable in the evaluation of the average of information in the system. The oscillation pseudo period of entropies in RbCl and CsI quantum pseudo dot increases suggesting an attenuation in the transmission of the information through the system while in the GaAs it is an inverse scenario. For the delta quantum dot, *(iii)* When the Gauss entropy is considered, a lot of information about the system is missed. The collapse revival phenomenon in Shannon entropy is observed in RbCl and GaAs delta quantum dot with the enhancement of delta parameter;  with the increase of this parameter,the system in the case of CsI evolve coherently; with Shannon and Tsallis entropies , information in the system is faster and coherently exchanged. *(iv)* The Shannon entropy is more significant because its amplitude outweighs on the others when $a$ is enhanced. The Tsallis entropy involves as wave bundle; which oscillate periodically with the increase of the oscillation period when $a$ is improved. Using the entropy of Tsallis, one can get the same information about the system as the entropy of Shannon; the only difference is that the latter is more significant in terms of volume of the information carried.